\documentclass[reprint,amssymb,amsmath,aip,cha]{revtex4-1}
\usepackage{docs}
\usepackage{bm}
\usepackage[colorlinks=true,linkcolor=blue,filecolor=blue,urlcolor=blue]{hyperref}%
\expandafter\ifx\csname package@font\endcsname\relax\else
\expandafter\expandafter
\expandafter\usepackage
\expandafter\expandafter
\expandafter{\csname package@font\endcsname}
\fi
\usepackage{color,graphicx}
\graphicspath{ {images/} }
\begin{document}

\title{Magneto-transport properties of proposed triply degenerate topological semimetal Pd$_{3}$Bi$_{2}$S$_{2}$}

\author{Shubhankar Roy, Arnab Pariari, Ratnadwip Singha, Biswarup Satpati, and Prabhat Mandal}

\affiliation{Saha Institute of Nuclear Physics, HBNI, 1/AF Bidhannagar, Calcutta 700 064, India}

\begin{abstract}
We report transport properties of single-crystalline Pd$_{3}$Bi$_{2}$S$_{2}$, which has been predicted to host an unconventional electronic phase of matter beyond three-dimensional Dirac and Weyl semimetals. Similar to several topological systems, the resistivity shows field-induced metal to semiconductor-like crossover at low temperature. Large, anisotropic and non-saturating magnetoresistance has been observed in transverse experimental configuration. At 2 K and 9 T, the MR value reaches as high as $\sim$1.1$\times$10$^{3}$ \%. Hall resistivity reveals the presence of two types of charge carriers and has been analyzed using two-band model. In spite of the large density ($>$ 10$^{21}$ cm$^{-3}$), the mobility of charge carriers is found to be quite high ($\sim$ 0.75$\times$10$^{4}$ cm$^{2}$ V$^{-1}$ s$^{-1}$ for hole and $\sim$ 0.3$\times$10$^{4}$ cm$^{2}$ V$^{-1}$ s$^{-1}$ for electron). The observed magneto-electrical properties indicate that Pd$_{3}$Bi$_{2}$S$_{2}$ may be a new member of the topological semimetal family, which can have a significant impact in technological applications.
\end{abstract}

\maketitle

Topology protected electronic properties of materials have opened up a new arena of research in modern condensed matter physics \cite{Hasan,XL,liu,Xu,RM,VMourik,SNadj}. These novel materials have generated immense research interest in fundamental physics of low-energy relativistic particles. In high-energy  physics, the relativistic fermions are protected by Poincare symmetry, while in condensed matter, they respect one of the 230 space group symmetries (SGs) of the crystal. The variation of crystal symmetry from one material to another escalates the potential to explore free fermionic excitations such as Dirac, Weyl, Majorana and beyond. Besides fundamental interest, the discovery of topological insulators \cite{Hasan,XL}, 3D Dirac and Weyl semimetals \cite{Xu,liu}, and Majorana fermions in superconducting heterostructures \cite{RM,VMourik,SNadj} have planted the seed of technological revolution in research and development of electronic devices. Fabrication of fast electronic devices, spintronics applications and fault tolerant quantum computing are the few among the several specific topics of interest. So, the prediction and experimental realization of new materials, which host such quasi-particle excitations, can have significant impact on paradigm shifting in technological application.\\

Very recently, Bradlyn \emph{et al.}\cite{Barry} have predicted the existence of exotic fermions near the Fermi level in several materials, governed by their respective space group symmetry. Unlike two- and four-fold degeneracy in 3D Weyl/Dirac semimetals, these systems exhibit three-, six-, and eight-fold degenerate band crossing at high symmetry points in the Brillouin zone. It has been proposed that Pd$_{3}$Bi$_{2}$S$_{2}$ with space group 199 hosts three-fold degenerate fermion as the quasi-particle excitation due to the three-band crossing at the high symmetry P point, which is just 0.1 eV above the Fermi level \cite{Barry}. So far, no magnetotransport experiment has been done on this material. The results are important not only for the fundamental research but also to manipulate them in technological application. In the present work, we have synthesised the single crystal of Pd$_{3}$Bi$_{2}$S$_{2}$ and performed the magnetotransport experiments in field up to 9 T and temperature down to 2 K.  \\
\begin{figure}
\includegraphics[width=0.5\textwidth]{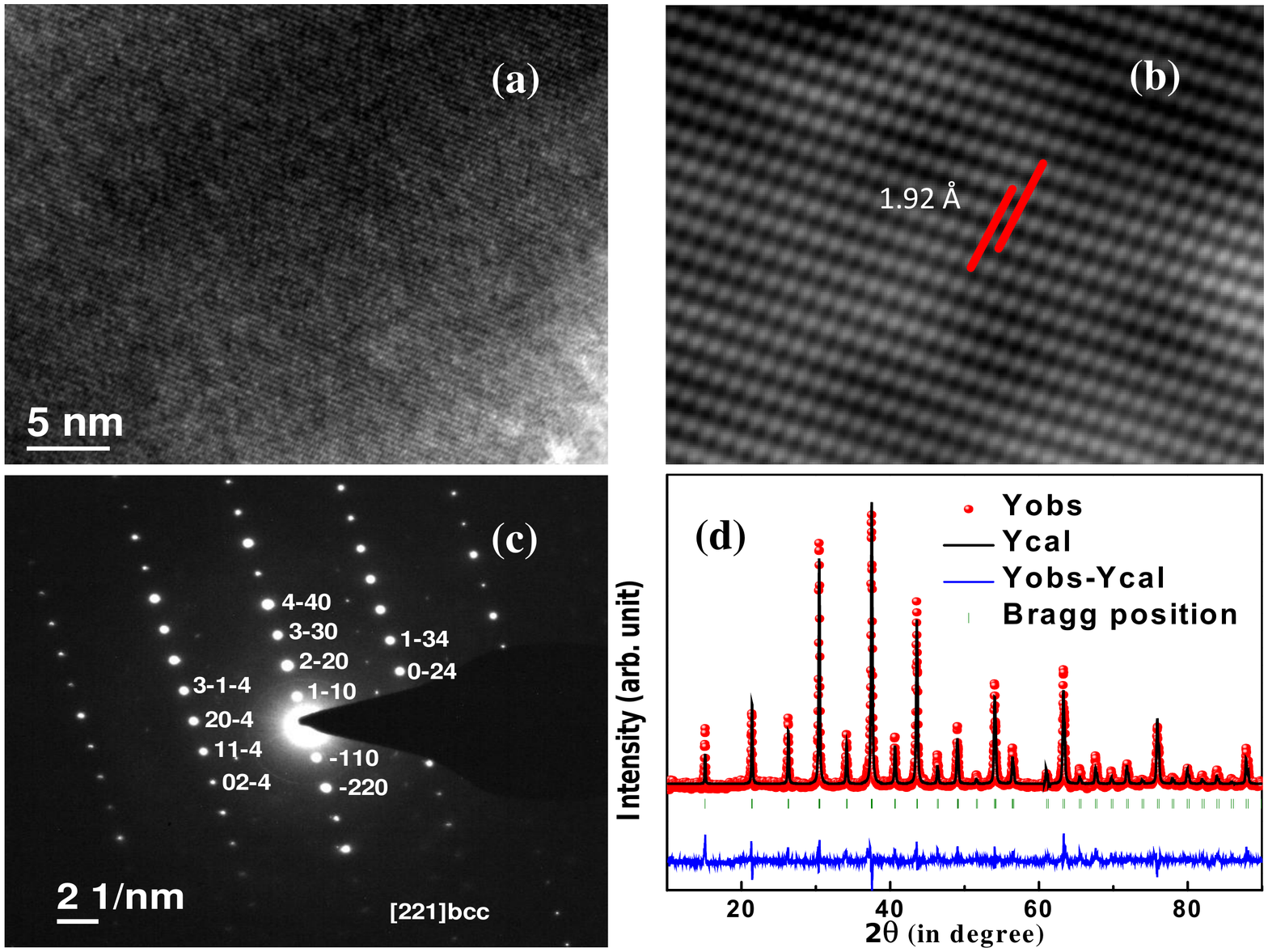}
\caption{(a) High resolution transmission electron microscopy image, taken on a representative piece of Pd$_{3}$Bi$_{2}$S$_{2}$ single crystal. (b) Fourier-filtered image of a small selected region in Fig. (a). (c) The selected area electron diffraction (SAD) patterns taken along [221]. (d) X-ray diffraction pattern of powdered single crystals of Pd$_{3}$Bi$_{2}$S$_{2}$. Red open circles are experimental data (Y$_{obs}$), black line is the calculated pattern (Y$_{cal}$), blue line is the difference between experimental and calculated intensities (Y$_{obs}$-Y$_{cal}$), and green lines show the Bragg positions.}
\label{f:1}
\end{figure}

Single crystal of Pd$_{3}$Bi$_{2}$S$_{2}$ was synthesized via standard self-flux method. At first, polycrystalline Pd$_{3}$Bi$_{2}$ was prepared by melting the stoichiometric mixture of high purity Pd pieces (Alfa Aesar 99.99\%) and Bi shot (Alfa Aesar 99.999\%) in an arc furnace, under argon atmosphere. Next, Pd$_{3}$Bi$_{2}$ and stoichiometric amount of sulphur pieces were kept at 350$^{\circ}$C for 24 h and at 900$^{\circ}$C for 48 h in a vacuum sealed quartz tube. Finally, the quartz tube was slowly cooled down to room temperature at the rate of 4$^{\circ}$C/h. Shiny plate-like crystals of typical dimensions $2.0\times0.4\times0.3$ mm$^{3}$ were mechanically extracted. Typical morphology and different crystallographic directions of a representative single crystal of Pd$_{3}$Bi$_{2}$S$_{2}$ have been shown in Fig. S1. Phase purity and the structural analysis of the samples were done using both high resolution transmission electron microscopy (HRTEM) in a FEI, Tecnai G$^{2}$ F30, S-Twin microscope equipped with energy dispersive x-ray spectroscopy (EDX, EDAX Inc.) unit and the powder x-ray diffraction (XRD) technique with Cu-K$_{\alpha}$ radiation in a Rigaku x-ray diffractometer (TTRAX III). The transport measurements have been performed in a 9 T physical property measurement system (Quantum Design) using ac-transport option and 10 mA current. The I-V characteristic of the sample has been recorded with current up to 40 mA to ensure that the electrical contacts are Ohmic. For both the resistivity and Hall measurements, the electrical contacts were made in four-probe configuration using conducting silver paint and gold wires (diameter 50 $\mu$m). The HRTEM image of a typical Pd$_{3}$Bi$_{2}$S$_{2}$ single crystal is shown in Figs. 1(a) and (b). Very clear periodic lattice structure implies that there is no secondary phase, atom clustering or disorder in the studied sample. Figure 1(c) shows the selected area electron diffraction (SAD) pattern recorded along [221] zone axis. The periodic pattern of the spots in SAD implies high-quality single crystalline nature of the grown samples. The diffraction pattern was indexed using the lattice parameters of cubic unit cell. To determine the stoichiometry, the EDX has been performed on different regions of the single crystal. In Fig. S2, we have shown one typical EDX spectrum. From the experimental results, the atomic ratio of Pd, Bi and S is found to be $\sim$2.93:2.28:1.79 throughout the sample with maximum experimental error $\pm$5 \%. However, for low $Z$ element sulphur, the statistical error in the EDX spectra is expected to be little higher. Figure 1(d) shows the high-resolution x-ray diffraction pattern of the powdered sample at room temperature. Within the resolution of XRD, we did not see any peak due to any impurity phase. Using the Rietveld profile refinement, we have calculated the lattice parameters $a$$ = $$b$$ = $$c$$ = $8.310(2) {\AA} with space group symmetry $I2_13$.\\

\begin{figure}
\includegraphics[width=0.5\textwidth]{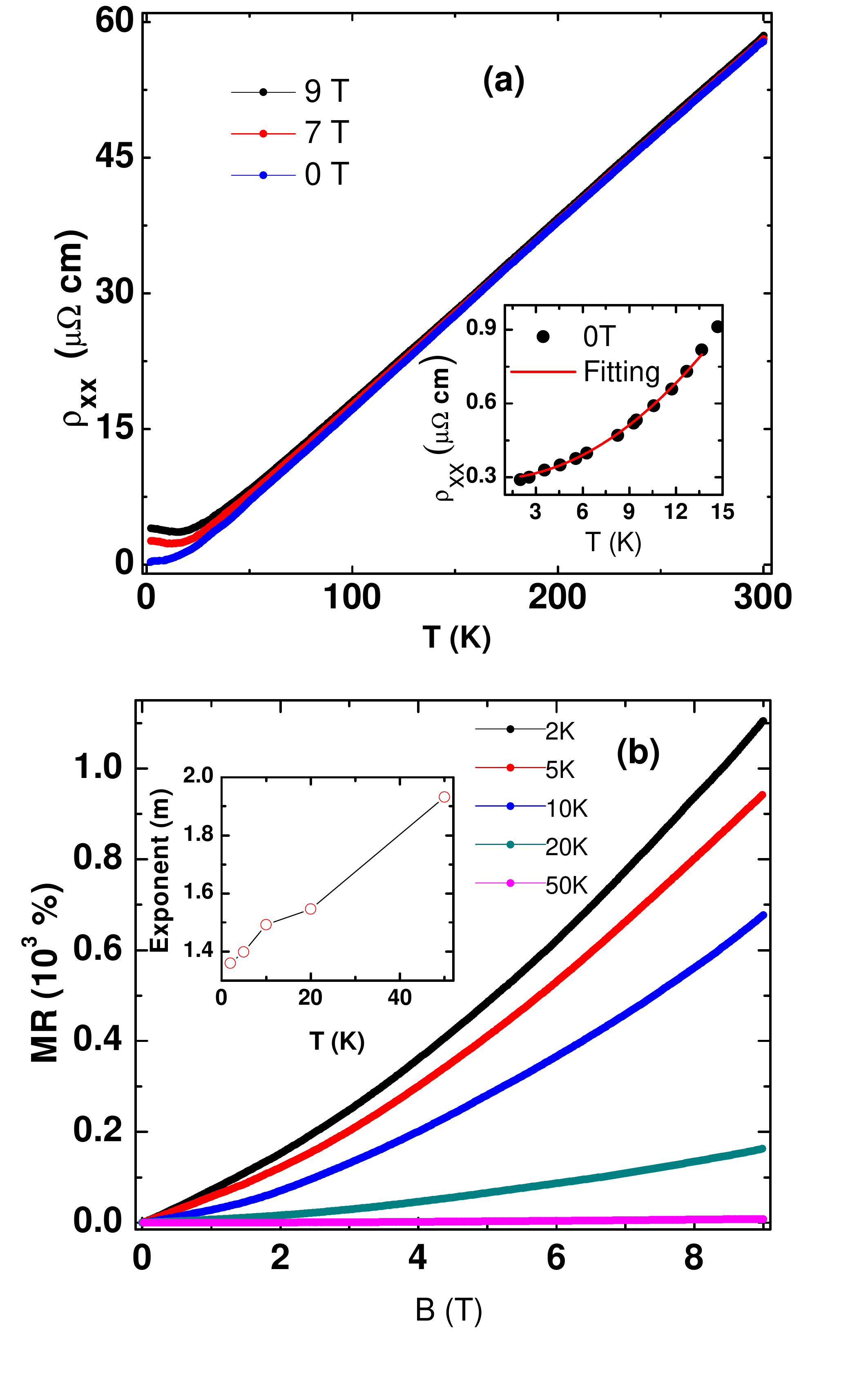}
\caption{(a) Temperature dependence of resistivity ($\rho_{xx}$) both in presence and absence of external magnetic field. Current ($I$) is applied perpendicular to magnetic field ($B$). Inset shows the low-temperature region fitted with $\rho_{xx}(T)=a+bT^2$. (b) Transverse magnetoresistance as a function of magnetic field at some representative temperatures. The current is applied along the crystallographic \textbf{a}-axis and the magnetic field is along \textbf{c}-axis. Inset shows the exponent ($m$) as a function of temperature, where MR at various temperatures have been fitted with the power law behavior, MR$\propto$$B^{m}$.}
\label{f:2}
\end{figure}

In Fig. \ref{f:2}(a), the zero-field resistivity ($\rho_{xx}$) of Pd$_{3}$Bi$_{2}$S$_{2}$ crystal shows metallic behavior over the whole range of temperature. $\rho_{xx}$ exhibits a monotonic decrement with the decrease in $T$ and becomes very small ($\sim$ 270 $n$$\Omega$ cm) at 2 K. The residual resistivity ratio (RRR), $\rho_{xx}$(300 K)/$\rho_{xx}$(2 K), is 213, which signifies good metallicity and high quality of single crystal. This value of RRR is larger than that reported ($\sim$ 110) earlier \cite{Masashi}. RRR of Pd$_{3}$Bi$_{2}$S$_{2}$ is comparable to that reported for several topological semimetals and metals \cite{Rsingha,Zhujun,Shekhar}. $\rho_{xx}$ up to $\sim$13 K can be fitted well with the equation, $\rho_{xx}(T)=a+bT^{2}$ [inset of Fig. \ref{f:2}(a)]. This indicates pure electronic correlation dominated scattering in the low-temperature region \cite{Ziman}. The behaviour of $\rho_{xx}(T)$ curve at low temperature is quite different from that reported for most of the Dirac and Weyl semimetals \cite{Rsingha,Shekhar,Tafti,Yue,Sun,RSingha2}, where $\rho_{xx}$ is found to be either very weakly dependent on $T$ or $\rho_{xx}$ $\propto$ $T^{n}$ with $n$$\geq$3. A $T^{2}$ dependence of $\rho_{xx}$ has only been observed in type-II Weyl semimetal WTe$_{2}$ \cite{Ylwang}. Almost linear $T$ dependence of $\rho_{xx}$ above $\sim$ 25 K suggests electron-phonon scattering dominated transport at high temperature. $T$ dependence of $\rho_{xx}$ for Pd$_{3}$Bi$_{2}$S$_{2}$ crystal has a striking similarity with conventional metal. With application of magnetic field, a small upturn in $\rho_{xx}(T)$ curve is seen in the low-temperature region. The field-induced metal to semiconductor-like crossover and low-temperature saturation-like behavior in resistivity are the commonly observed phenomena in topological semimetals \cite{Shekhar,Tafti,Yue,Sun}. \\

The magnetoresistance of Pd$_{3}$Bi$_{2}$S$_{2}$ in transverse experimental configuration ($I\parallel$\textbf{a}-axis, $B\parallel$\textbf{c}-axis) has been measured at various temperatures. As shown in Fig. 2(b), MR, which is defined as [$\rho_{xx}(B)$-$\rho_{xx}(0)$]/$\rho_{xx}(0)$, is positive and large at low temperature and shows superlinear field dependence up to 9 T. At 2 K and 9 T, the value of MR is about 1.1$\times$10$^{3}$ \%, which is comparable to that observed in several Dirac and Weyl materials \cite{Arnab2,Orest,Yong,Aifeng,Silu,Max}. With the increase in temperature, however, MR decreases rapidly. MR at various temperatures follows the power-law behavior, MR$\propto$ $B^{m}$. Unlike conventional metals and topological semimetals, exponent ($m$) is found to be extremely sensitive to temperature. From the inset of Fig. 2(b), one can see that $m$ increases almost linearly with the increase in temperature. As a result of such unusual behavior of MR, strong violation of the Kohler scaling is observed, which is shown in Fig. 3(a). This violation is ascribed to various aspects of charge conduction mechanism such as the presence of more than one types of charge carrier and different temperature dependence of their mobilities ($\mu$) \cite{Ylwang,RSingha2,Mcken,LWang}.\\
\begin{figure}
\includegraphics[width=0.5\textwidth]{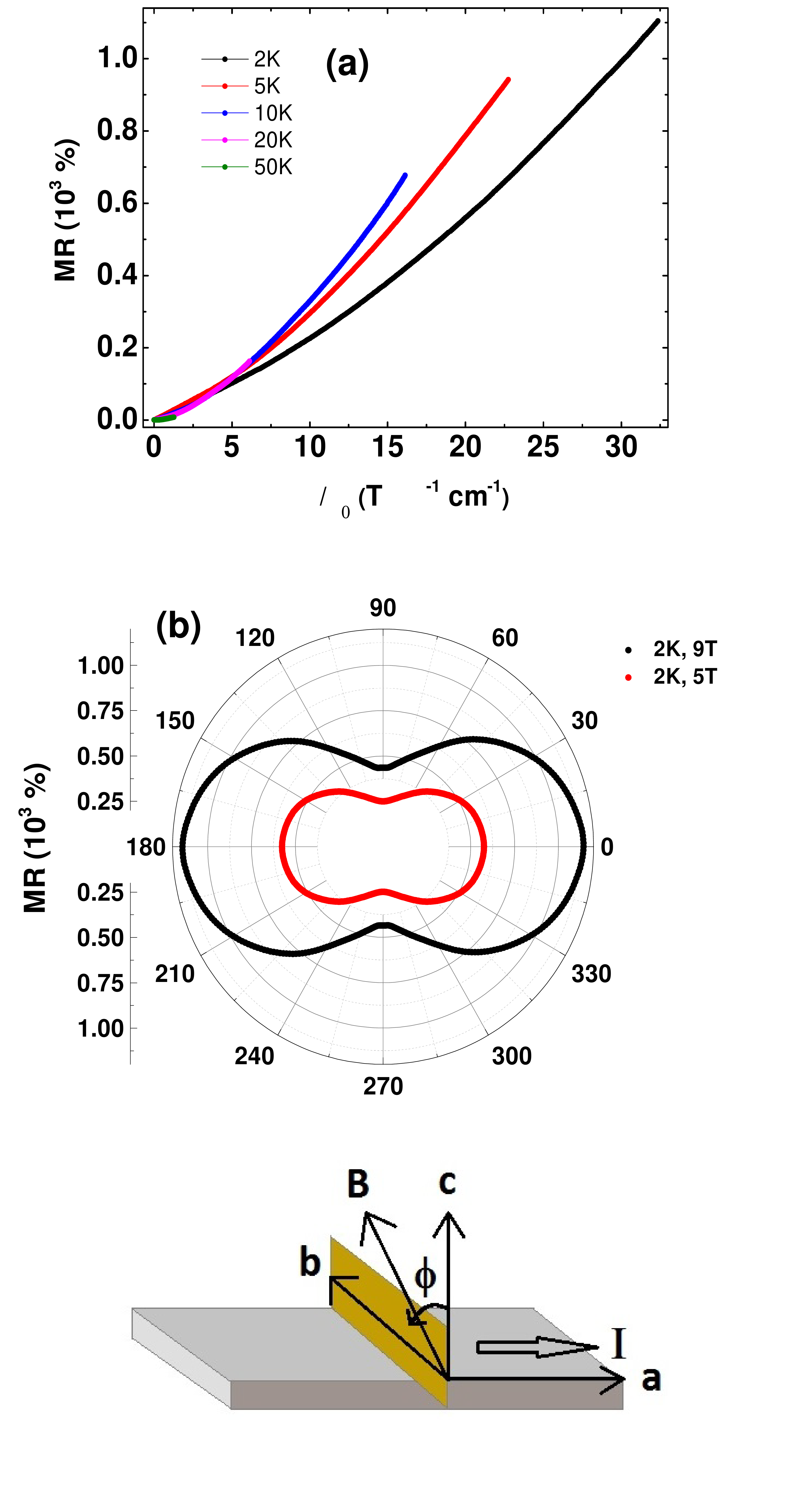}
\caption{(a) MR at different temperatures are scaled using Kohler's rule. (b) Angular dependence of MR for 2 K and magnetic field 5 and 9 T. The magnetic field is rotated about the crystallographic \textbf{a}-axis. }
\end{figure}

With the help of a sample rotator, we have studied the angular dependence of MR. In this experimental set up, the current direction has been kept fixed along crystallographic \textbf{a}-axis on the largest flat plane, and the magnetic field is rotated about \textbf{a}-axis, as illustrated in the schematic diagram of  Fig. 3(b). The anisotropy in MR is evident from the experimental data at 2 K, which has been plotted in Fig. 3(b) for two representative magnetic field strengths 5 and 9 T. MR is maximum when the magnetic field is perpendicular to the plane of the sample ($\phi$$=$0) and it decreases as the field is rotated towards inplane direction and becomes minimum at $\phi$=90$^{\circ}$. The anisotropy ratio in MR is $\sim$ 2.2. Similar anisotropic behavior of MR has been reported in several topological semimetals \cite{Zhujun,Zhao,Petrovic}. In a non-magnetic material, MR is mainly determined by the  mobility of the charge carrier in the plane perpendicular to the external magnetic field \cite{Aure}. Again, the mobility  is defined as the ratio of the scattering time ($\tau$) to the effective mass of the charge carrier (m$_{eff}$), $\mu$ $\sim$ $\frac{\tau}{m_{eff}}$. Both the quantities $\tau$, which depends on the cross-sectional area of the Fermi surface in the plane perpendicular to magnetic field, and m$_{eff}$, which depends on the electronic band curvature, are directly related to Fermi surface. As a consequence, the nature and geometry of the Fermi surface play a crucial role in determining the angle dependence of transverse magnetoresistance. The anisotropy in MR can be explained by considering the anisotropy in the Fermi surface of Pd$_{3}$Bi$_{2}$S$_{2}$. In a material with simple cubic crystal structure, one should not expect any anisotropy in the Fermi surface and as a consequence, the angular dependence in MR will be negligible. However, the cubic crystal structure of Pd$_{3}$Bi$_{2}$S$_{2}$ has chiral nature $-$ absence of center of inversion and mirror planes \cite{Masashi}. Due to this lower crystal symmetry, one expects anisotropic Fermi surface and angular dependence in the orbital magnetoresistance induced by the broken four-fold symmetry in cubic structure. Often, topological semimetals show negative longitudinal magnetoresistance (LMR), where current is parallel to the magnetic field \cite{Xhuang,huili,quing,Pariari}. We have not observed any negative LMR in the present compound. On the contrary, we have observed a positive LMR, which has been shown and discussed in supplementary material (see Fig. S3).\\
\begin{figure}
\includegraphics[width=0.48\textwidth]{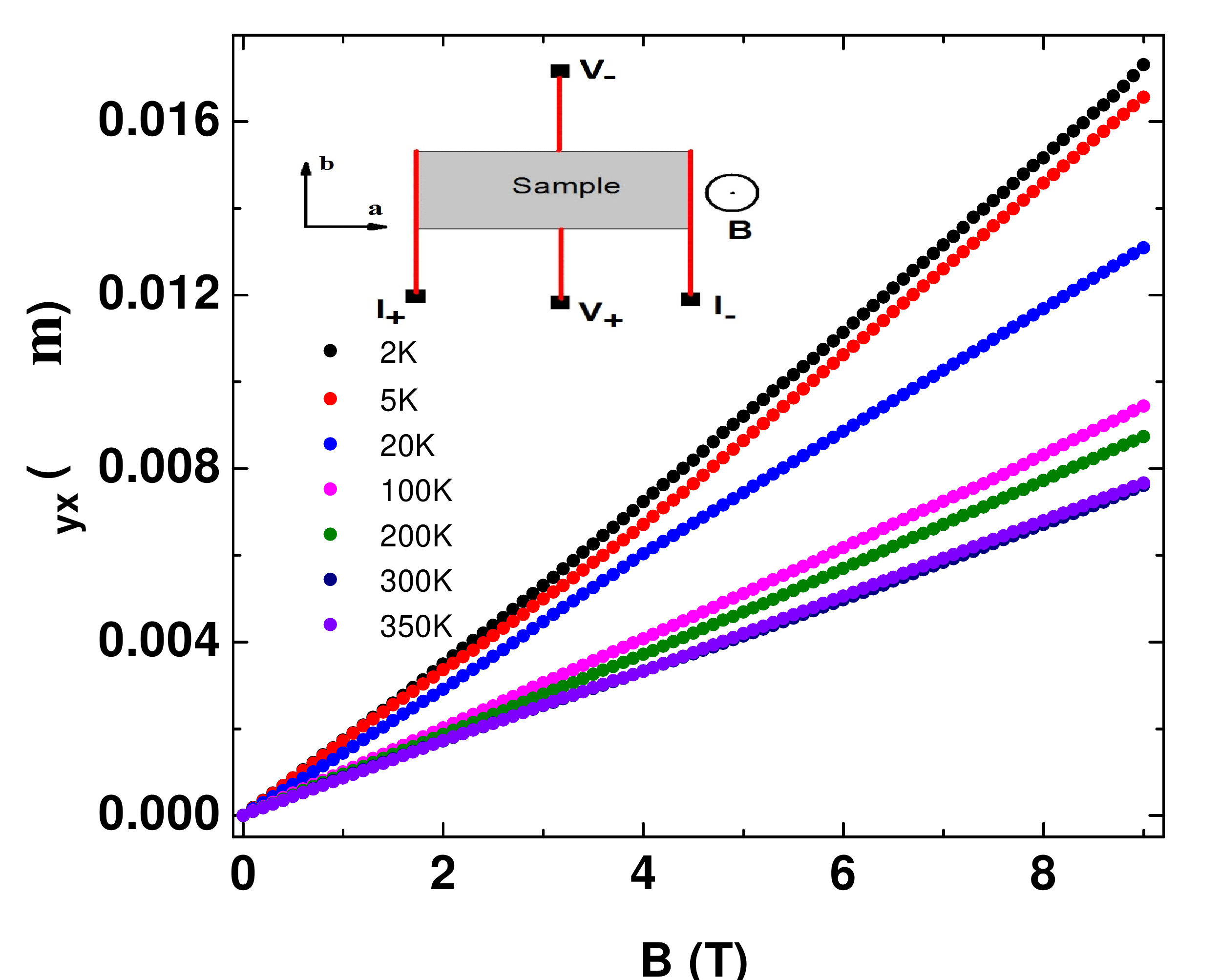}
\caption{The field dependence of Hall resistivity ($\rho$$_{yx}$) at different temperatures. A schematic of Hall measurement set up is shown in inset.}
\end{figure}

To determine the type of charge carrier and its mobility, the Hall resistivity ($\rho$$_{yx}$) measurement has been performed. $\rho$$_{yx}$ has been obtained by the operation, $\rho$$_{yx}$= [$\rho$$_{yx}$ (positive field)-$\rho$$_{yx}$ (negative field)]/2, to remove the resistivity contribution induced by the small unavoidable mismatch in Hall resistivity contacts on the opposite side of the sample. In Fig. 4, the field dependence of $\rho$$_{yx}$ is shown at some representative temperatures. At high temperature, $\rho$$_{yx}$ is almost linear in field and positive. However, with decreasing temperature, $\rho$$_{yx}$($B$) becomes weakly non-linear. This implies the presence of more than one types of charge carrier. In presence of external magnetic field, the electrical conductivity ($\sigma_{xx}$) and Hall conductivity ($\sigma_{xy}$) are given by the expressions, $\sigma_{xx}$=$\frac{\rho_{xx}}{\rho_{yx}^{2}+\rho_{xx}^{2}}$ and $\sigma_{xy}$=$\frac{\rho_{yx}}{\rho_{yx}^{2}+\rho_{xx}^{2}}$, following the tensorial inversion of the resistivity matrix. In Fig. 5(a), we have plotted $\sigma_{xy}$ as a function of magnetic field for a representative temperature 2 K. The Hall conductivity shows a non-monotonic behavior, which can be explained from the field dependence of $\rho_{yx}$ and $\rho_{xx}$. In absence of magnetic field, as $\rho$$_{yx}$=0 and $\rho$$_{xx}$$\neq$0, $\sigma_{xy}$ is zero. In low field region, $\sigma_{xy}$ is small and increases with increasing magnetic field. With further increase in field strength, $\rho$$_{yx}$ increases monotonically but the denominator ($\rho_{yx}^{2}+\rho_{xx}^{2}$) becomes very large due to large enhancement of $\rho$$_{xx}$ from its zero field value. Hence, the field dependence of $\sigma_{xy}$ shows a low field maximum and tends towards zero at high fields. The field dependence of $\rho_{yx}$ and $\frac{1}{\rho_{yx}^{2}+\rho_{xx}^{2}}$ can make this discussion more clear, which have been shown in Fig S5. Considering the semiclassical two-band model, the following expressions of $\sigma_{xy}$ and $\sigma_{xx}$ have been used to fit the experimental data \cite{Xhuang}.
\begin{equation}\label{1}
 \sigma_{xy}(B)=[n_{h}\mu_{h}^{2}\frac{1}{1+(\mu_{h}B)^{2}}-n_{e}\mu_{e}^{2}\frac{1}{1+(\mu_{e}B)^{2}}]eB
\end{equation}
\begin{equation}\label{2}
 \sigma_{xx}(B)=e[n_{h}\mu_{h}\frac{1}{1+(\mu_{h}B)^{2}}+n_{e}\mu_{e}\frac{1}{1+(\mu_{e}B)^{2}}]
\end{equation}
where n$_{h}$(n$_{e}$) and $\mu_{h}$($\mu_{e}$) are hole(electron) density and mobility, respectively. We have done the global fitting of $\sigma_{xy}(B)$ and $\sigma_{xx}(B)$ [where these two quantities have been fitted simultaneously using Eqs. (1) and (2)] for all temperatures and in Fig. 5(a), we have shown the global fitting at  2 K as a representative. As evident from Fig. 5(b), the mobilities show strong temperature dependence, while the density of the two types of charge carrier [inset of Fig. 5(b)] remains almost invariant throughout the temperature range within the accuracy of our two-band analysis. These results suggest considerable contribution from both types of charge carrier. Fig. S6 also clearly depicts that within the accuracy of our two-band analysis and experimental error, the ratio $\frac{n_{h}}{n_{e}}$ remains almost same. Whereas,  $\frac{\mu_{h}}{\mu_{e}}$  increases to a large value ($\sim11$) at higher temperatures from $\sim2.6$ at 2 K. As hole and electron density are comparable, mobility plays important role on charge conduction. For this reason, the low temperature $\rho$$_{yx}$ vs B plot appears to be weakly non-linear with upward curvature, compared to the  $\rho$$_{yx}$ vs B plot at higher temperatures, where hole-type carriers dominate the transport properties. The obtained hole (electron) density $\sim$ 1.8$\times$10$^{21}$ cm$^{-3}$ (1.6$\times$10$^{21}$ cm$^{-3}$) is quite large compared to that usually observed ($\sim$ 10$^{17}$-10$^{19}$ cm$^{-3}$) in Dirac and Weyl semimetals \cite{Tafti,Rsingha,Zhujun,Shekhar,RSingha2}. However, the carrier density in Pd$_{3}$Bi$_{2}$S$_{2}$ is comparable to that observed in topological nodal-line semimetals ($n$$\sim$10$^{20}$-10$^{23}$ cm$^{-3}$) \cite{Hu,Sankar} and conventional metals. In spite of large carrier density, the mobility of charge carrier in Pd$_{3}$Bi$_{2}$S$_{2}$ is found to be quite high ($\sim$ 0.75$\times$10$^{4}$ cm$^{2}$ V$^{-1}$ s$^{-1}$ for hole and $\sim$ 0.3$\times$10$^{4}$ cm$^{2}$ V$^{-1}$ s$^{-1}$ for electron) and comparable to that reported for different topological electronic materials. On the other hand, in conventional metal, the mobility of charge carrier is low and  the MR obeys the Kohler scaling. It appears from the above discussion that some of the properties of Pd$_{3}$Bi$_{2}$S$_{2}$ are similar to that of conventional metals, whereas some are similar to topological semimetals.\\
\begin{figure}
\includegraphics[width=0.48\textwidth]{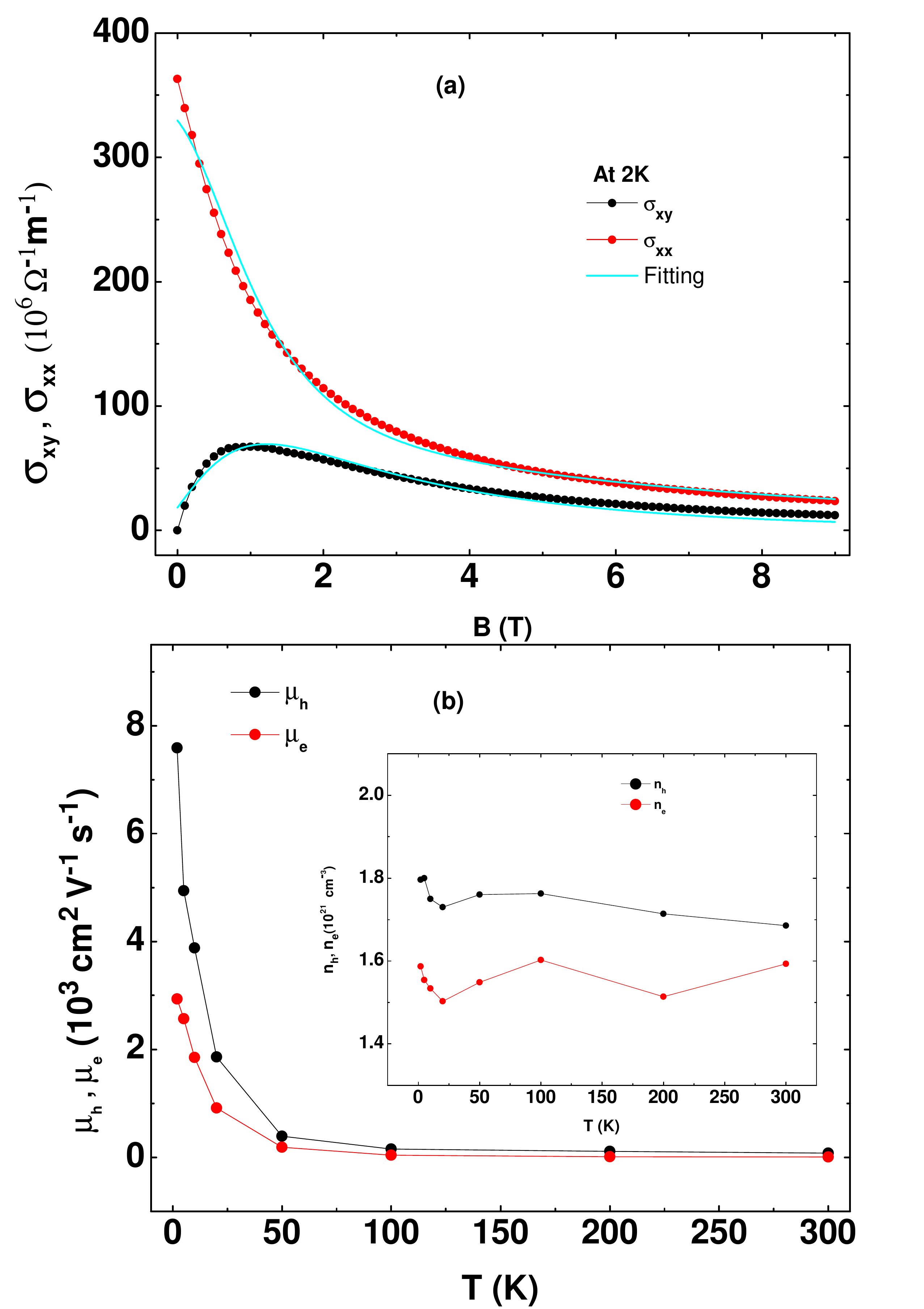}
\caption{ (a) Global fitting of the Hall conductivity ($\sigma_{xy}$) and conductivity ($\sigma_{xx}$) using two-band model [Eqs. (1) and (2)]. (b) Temperature dependence of electron and hole carrier mobility obtained from global fitting. In inset electron and hole carrier density have been shown as a function of temperature.}
\end{figure}

In conclusion, we have synthesized a newly discovered electronic material Pd$_{3}$Bi$_{2}$S$_{2}$ and performed the systematic magneto-electronic transport experiments. A large non-saturating transverse magnetoresistance has been observed. The large anisotropy ratio in MR indicates significant anisotropy in the Fermi surface. Hall measurement reveals high mobility of charge carriers ($\sim$ 0.75$\times$10$^{4}$ cm$^{2}$ V$^{-1}$ s$^{-1}$ for hole and $\sim$ 0.3$\times$10$^{4}$ cm$^{2}$ V$^{-1}$ s$^{-1}$ for electron), in spite of the unusually large carrier density ($>$ 10$^{21}$ cm$^{-3}$). Similar to conventional metal, $\rho$$_{xx}$ shows $T^{2}$ dependence at low temperature and $T$ dependence at high temperature. However, MR does not obey Kohler scaling. The present results suggest that Pd$_{3}$Bi$_{2}$S$_{2}$ may host unconventional electronic band structure beyond three-dimensional Dirac/Weyl semimetals and can be a possible candidate for several technological applications.

\begin{center}
\textbf{Supplementary Material}\\
\end{center}
In the supplementary material, we have shown the crystal structure and image of a typical single crystal. Detailed information regarding the EDX, longitudinal magnetoresistance and scaling of magnetoresistance have been provided. We have also included additional discussions and plots regarding the field dependence of Hall conductivity and mobility ratio of the two types of carriers, in this section.

\end{document}